\documentclass[twocolumn,showpacs,amsmath,amssymb]{revtex4}
\usepackage{graphicx}% Include figure files
\usepackage{array}% Align table columns on decimal point
\usepackage{bm}% bold math

\newcommand{\be}{\begin{equation}}
\newcommand{\ee}{\end{equation}}
\newcommand{\ben}{\begin{eqnarray}}
\newcommand{\een}{\end{eqnarray}}

\newcommand{\iii}{\'{\i}}

\begin{document}

%\preprint{APS/123-QED}

\title{Properties of a geometric measure for quantum discord}

\author{J. Batle$^{1}$,  A. Plastino$^{2}$,  A.R. Plastino$^{3,\,4}$, M. Casas$^{1}$}
\affiliation{ $^1$Departament de F\iii sica and IFISC, Universitat de les
Illes Balears,
 07122 Palma de Mallorca, Spain  \\\\
$^2$IFLP-CCT-CONICET, National University La Plata,
  C.C. 727, 1900 La Plata, Argentina  \\\\ $^3$CREG-UNLP-CONICET,
 National University la Plata, C.C. 727, 1900 La Plata, Argentina
\\\\ $^4$Instituto Carlos I de Fisica Teorica
 y Computacional, Universidad de Granada, Granada, Spain}

\date{\today}% It is always \today, today,
             %  but any date may be explicitly specified

\begin{abstract}
We discuss some properties of the quantum discord based on the
geometric measure advanced by Dakic, Vedral, and Brukner [Phys.
Rev. Lett. {\bf 105}, 190502 (2010)], with emphasis on Werner- and
MEM-states. By recourse to a systematic survey of the two-qubits state-space
we ascertain just how good the measure is in representing quantum discord.
We explore the dependence of quantum discord on the degree of
mixedness of the bipartite states, and also its connection with non-locality
as measured by the maximum violation of a Bell inequality within the CHSH
scenario.
\end{abstract}

% PACS, the Physics and Astronomy
                             % Classification Scheme.
%\keywords{Suggested keywords}%Use showkeys class option if keyword
                              %display desired
\maketitle

\section{\label{sec:intro}Introduction}

One of the most fundamental concepts in the  quantum description
of Nature is that of entanglement, that in recent years has been
the subject of intense research efforts \cite{BZ06,HHHH09,AFOV08,NC00}.
All entangled pure states of bipartite systems exhibit non-local
features that manifest themselves through the violation of Bell
inequalities \cite{NC00,P93}. On the other hand, there exist entangled mixed states
that comply with all Bell inequalities. Therefore, entanglement
encompasses a concept of quantum correlations broader than the one
associated with non-locality. Entanglement, however, does not describe
all aspects of the quantum correlations exhibited by multipartite physical
systems. In this regard, an important concept that has been receiving much
attention lately is that of quantum discord, that refers to
quatum correlations different from those involved in entanglement
\cite{geom,olli,zambrini,ferraro,dattaprl,luo}. Besides its intrinsic
conceptual interest, the study of quantum discord may also have
technological implications: examples of improved quantum computing
tasks that take advantage of quantum correlations but do not rely
on entanglement have been reported [see for instance, among a
quite extensive references-list,
\cite{geom,olli,zambrini,ferraro,dattaprl,luo}]. The aim of the present
contribution is to investigate various features of a recently
advanced geometric measure of quantum discord and its relationships
with the degree of mixture of the quantum state describing a multi-partite system. T
The connection between discord and non-locality will also be addressed.

Quantum discord \cite{olli} constitutes a quantitative measure of
the ``non-classicality" of bipartite correlations as given by the
discrepancy between the quantum counterparts of two classically equivalent
expressions for the mutual information. More precisely, quantum discord is
defined as the difference between two ways of expressing (quantum mechanically)
such an important entropic quantifier. If $S$ stands for the von Neumann entropy,
for a bipartite state $A-B$ of density matrix $\rho$ and reduced
(``marginals") ones $\rho_A$-$\rho_B$, the quantum mutual
information (QMI) $M_q$ reads \cite{olli}

\be \label{uno} M_q(\rho)= S(\rho_A) +  S(\rho_B) - S(\rho), \ee
which is to be compared to its associated classical notion
$M_{class}(\rho)$, that is expressed using conditional entropies.
If a complete projective measurement $\Pi_j^B$ is performed on B
and (i) $p_i$ stands for $Tr_{AB}\,\Pi_i^B\,\rho$ and (ii)
$\rho_{A|\vert \Pi_i^B}$ for $[\Pi_i^B\,\rho\,\Pi_i^B/p_i]$, then
our conditional entropy becomes

\be  \label{unobis}  S(A \vert\,\{ \Pi_j^B \}) =\sum_i\,p_i\,
S(\rho_{A|\vert \Pi_i^B}), \ee so that $M_{class}(\rho)$ adopts
the appearance

\be \label{dos} M_{class}(\rho)_{\{ \Pi_j^B \}} = S(\rho_A)- S(A
\vert\,\{ \Pi_j^B \}). \ee Now, if we minimize over all possible $
\Pi_j^B$ the difference $M_q(\rho)-M_{class}(\rho)_{\{ \Pi_j^B
\}}$ we obtain the quantum discord $\Delta$, that quantifies
non-classical correlations in a quantum system, {\it  including
those not captured by entanglement}. One notes then that only
states with zero $\Delta$ may exhibit strictly classical
correlations. An interesting explorative work is that of Zambrini
et al. \cite{zambrini} that illustrate on several interesting
aspects of quantum discord
 for a large family of states. A. Ferraro et al. \cite{ferraro} have shown that such
kind of  these states has negligible Hilbert space (HS) volume. In
other words, a state picked out at random from HS must exhibit
positive discord. In this reference, a simple necessary criterion
for zero quantum discord is also given. Intuitively, quantum
discord may be viewed as the minimal correlations' loss (as
measured by the quantum mutual information) due to measurement, an
 interpretation  entirely analogous to the original
one of  Ollivier and Zurek
\cite{geom,olli,zambrini,ferraro,dattaprl,luo}.

Despite increasing evidences for relevance of the quantum discord
(Qd) in describing non-classical resources in information
processing tasks, there was until quite recently no
straightforward criterion to verify the presence of discord in a
given quantum state. Since its evaluation involves an optimization
procedure and analytical results are known only in a few cases,
such criteria become clearly desirable. last year Datta
advanced a condition for nullity of quantum discord \cite{datta},
and progress was also achieved in \cite{geom} by introducing an
interesting geometric measure of quantum discord (GQD).
Let $\chi$ be a generic $\Delta=0-$state. The GQD measure is then
given by

\be \label{tres} D(\rho)= {\rm Min}_{\chi}[||\rho-\chi||^2],  \ee

\noindent
where the minimum is over the set of zero-discord states $\chi$.
We deal then with  the square of Hilbert-Schmidt norm of Hermitian
operators, $||\rho-\chi||^2= Tr[(\rho-\chi)^2]$ . Dakic et al.
show how to evaluate this quantity for an arbitrary two-qubit
state  \cite{geom,luo}.  Moreover, they demonstrate  the their
geometric distance contains all relevant information associated to
the notion of quantum discord. This was a remarkable feat given
that, despite robust evidence for the pertinence  of the
Qd-notion,
 its evaluation involves optimization
procedures, with analytical results being known only in a few
cases.

%%%%%%%%%%%%%%%%%%%%%%%%%%%%%%%%%%%%%%%%%%%%%%%%%%%%%%%%%%%%%%%%%%%%%%%%%%%%%%%%%%%%%%%%%%%%%%%%%%%%%%%%%%
%%%%%%%%%%%%%%%%%%%%%%%%%%%%%%%%%%%%%%%%%%%%%%%%%%%%%%%%%%%%%%%%%%%%%%%%%%%%%%%%%%%%%%%%%%%%%%%%%%%%%%%%%%
Now, given the general form of an arbitrary two-qubits' state
 in the Bloch representation
\ben \label{rhoBloch}& 4\rho=   \mathcal{I} \otimes \mathcal{I} +
\sum_{u=1}^{3} x_u \sigma_u \otimes \mathcal{I} + \sum_{u=1}^{3}
y_u \mathcal{I} \otimes \sigma_i + \cr & +\sum_{u,v=1}^{3} T_{uv}
\sigma_u \otimes \sigma_v, \een \noindent with $x_u=Tr(\rho
(\sigma_u \otimes \mathcal{I}))$, $y_u=Tr(\rho (\mathcal{I}
\otimes \sigma_u))$, and $T_{uv}=Tr(\rho (\sigma_u \otimes
\sigma_v))$, it is found in Ref. \cite{geom} that a necessary and
sufficient criterion for witnessing non-zero quantum discord is
given by the rank of the correlation matrix

\begin{equation} \label{Rmatrix}
\frac{1}{4} \left( \begin{array}{cccc}
1 & y_1 & y_2 & y_3\\
x_1 & T_{11} & T_{12} & T_{13}\\
x_2 & T_{21} & T_{22} & T_{23}\\
x_3 & T_{31} & T_{32} & T_{33}
\end{array} \right),
\end{equation}

\noindent that is, a state $\rho$ of the form (\ref{rhoBloch})
exhibits finite quantum discord iff the matrix (\ref{Rmatrix}) has
a rank greater that two. It is seen that the  geometric measure
(\ref{tres}) is of the final form \cite{geom}

\ben \label{Dfinal} & D(\rho)=\frac{1}{4} \bigg( ||{\bf x}||^2 +
|| T ||^2  - \lambda_{\max} \bigg)=\cr &=
\frac{1}{R}-\frac{1}{4}-\frac{1}{4}\bigg( ||{\bf y}||^2 +
\lambda_{\max} \bigg), \een \noindent where $||{\bf x}||^2=\sum_u
x_u^2$, and $\lambda_{\max}$ is the maximum eigenvalue of the
matrix $(x_1,x_2,x_3)^t (x_1,x_2,x_3) + TT^t$. Here the
superscript $t$ denotes either vector or matrix transposition. The
second expression emphasizes the natural dependence of $D$ on the
participation ratio $R=1/Tr(\rho^2)$. Notice that this measure is
intimately connected with the quantities appearing in
(\ref{Rmatrix}).
%%%%%%%%%%%%%%%%%%%%%%%%%%%%%%%%%%%%%%%%%%%%%%%%%%%%%%%%%%%%%%%%%%%%%%%%%%%%%%%%%%%%%%%%%%%%%%%%%%%%%%%%%%
%%%%%%%%%%%%%%%%%%%%%%%%%%%%%%%%%%%%%%%%%%%%%%%%%%%%%%%%%%%%%%%%%%%%%%%%%%%%%%%%%%%%%%%%%%%%%%%%%%%%%%%%%%
It is out goal in this communication that of applying the
techniques advanced in \cite{geom} to elucidate further, hopefully
interesting, Qd-facets, beginning by ascertaining just how well
the measure represents the quantum discord notion.

\section{Typical features of the geometrical measure of quantum discord (GQD)}
\subsection{Preliminaries}

We shall perform a systematic numerical survey of the properties
of arbitrary (pure and mixed) states of a given quantum system by
recourse to an exhaustive exploration of the concomitant
state-space ${\cal S}$. To such an end it is necessary to
introduce an appropriate measure $\mu $ on this space. Such a
measure is needed to compute volumes within ${\cal S}$, as well as
to determine what is to be understood by a uniform distribution of
states on ${\cal S}$.  The natural measure that we are going to
adopt here is taken from the work of Zyczkowski {\it et al.}
\cite{ZHS98,Z99}. An arbitrary (pure or mixed) state $\rho$ of a
quantum system described by an $N$-dimensional Hilbert space can
always be expressed as the product of three matrices,

\be \label{udot} \rho \, = \, U D[\{\lambda_i\}] U^{\dagger}. \ee

\noindent Here $U$ is an $N\times N$ unitary matrix and
$D[\{\lambda_i\}]$ is an $N\times N$ diagonal matrix whose
diagonal elements are $\{\lambda_1, \ldots, \lambda_N \}$, with $0
\le \lambda_i \le 1$, and $\sum_i \lambda_i = 1$. The group of
unitary matrices $U(N)$ is endowed with a unique, uniform measure:
the Haar measure $\nu$ \cite{PZK98}. On the other hand, the
$N$-simplex $\Delta$, consisting of all the real $N$-uples
$\{\lambda_1, \ldots, \lambda_N \}$ appearing in (\ref{udot}), is
a subset of a $(N-1)$-dimensional hyperplane of ${\cal R}^N$.
Consequently, the standard normalized Lebesgue measure ${\cal
L}_{N-1}$ on ${\cal R}^{N-1}$ provides a natural measure for
$\Delta$. The aforementioned measures on $U(N)$ and $\Delta$ lead
then to a natural measure $\mu $ on the set ${\cal S}$ of all the
states of our quantum system \cite{ZHS98,Z99,PZK98,BCPP02}, namely,

\be \label{memu1}
 \mu = \nu \times {\cal L}_{N-1}.
 \ee

 \noindent
  All our present considerations are based on the assumption
 that the uniform distribution of states of a quantum system
 is the one determined by the measure (\ref{memu1}). Thus, in our
 numerical computations we are going to randomly generate
 states according to the measure (\ref{memu1}).

\subsection{Probability density distributions associated with $D[\rho]$}

Let us emphasize  that quantum discord quantifies nonclassical
correlations in a quantum system and one is most interested in
those  not captured by entanglement. Only states with zero discord
exhibit strictly classical correlations. As stated above, Ferraro
et al. \cite{ferraro} proved that these number of such states is
``negligible" for the whole of Hilbert's space (HS), so that  a
quantum state picked up at random has positive discord, a result
that holds for any HS-dimension and has straightforward
implications for quantum computation.

Maximally-entangled mixed states (MEMS) were first studied by
Munro et al. \cite{MJWK01} and are of special interest for quantum
communication purposes. For them, the quantum discord can be
easily obtained. Their connection with quantum discord has been
recently studied by Zambrini et al. \cite{zambrini}. We will find
these states below.

\begin{figure}[htbp]
\begin{center}
\includegraphics[width=8cm]{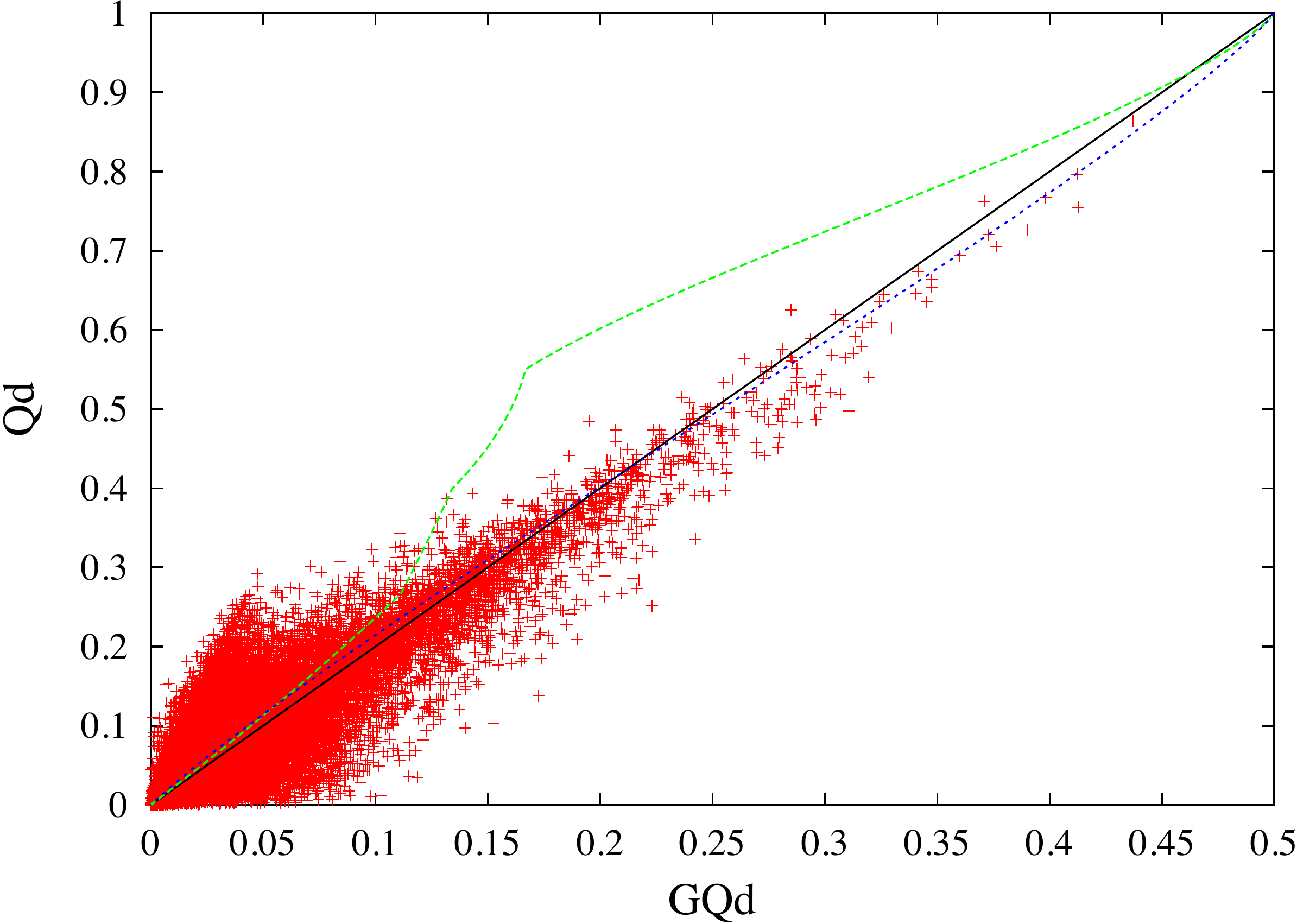}
\caption{(Color online) Sample plot of discord measures Qd vs GQd (crosses) for two qubit
mixed sates. There exists a strong correlation between both quantities, the plane being mostly
populated for low values of the either quantum discord measures. The upper dashed line represents
the MEMS states. The diagonal (solid line) is closely followed by the curve (dotted line) representing
Werner states.}
\label{fig1}
\end{center}
\end{figure}

We ask ourselves first of all for the question of how well does
the values of the distance $D$ represent the presence of quantum
discord. For the answer we refer the reader to Fig. 1, which plots
the Qd-amount versus the geometric Qd values $D$ of states picked
at random from the 2-qubits space. Absolute correspondence would
be represented by the diagonal at 45 degrees. Crosses represent Qd
and GQd for an arbitrary state. We see that there is a good
correlation between the two quantities. The upper  dashed line
represents the MEMS states (see below). The diagonal (solid line)
is closely followed by the curve (dotted line) representing Werner
states. We have found no other states correlated with Qd and GQd
to a greater extent than those of these two kinds.

\begin{figure}[htbp]
\begin{center}
\includegraphics[width=8cm]{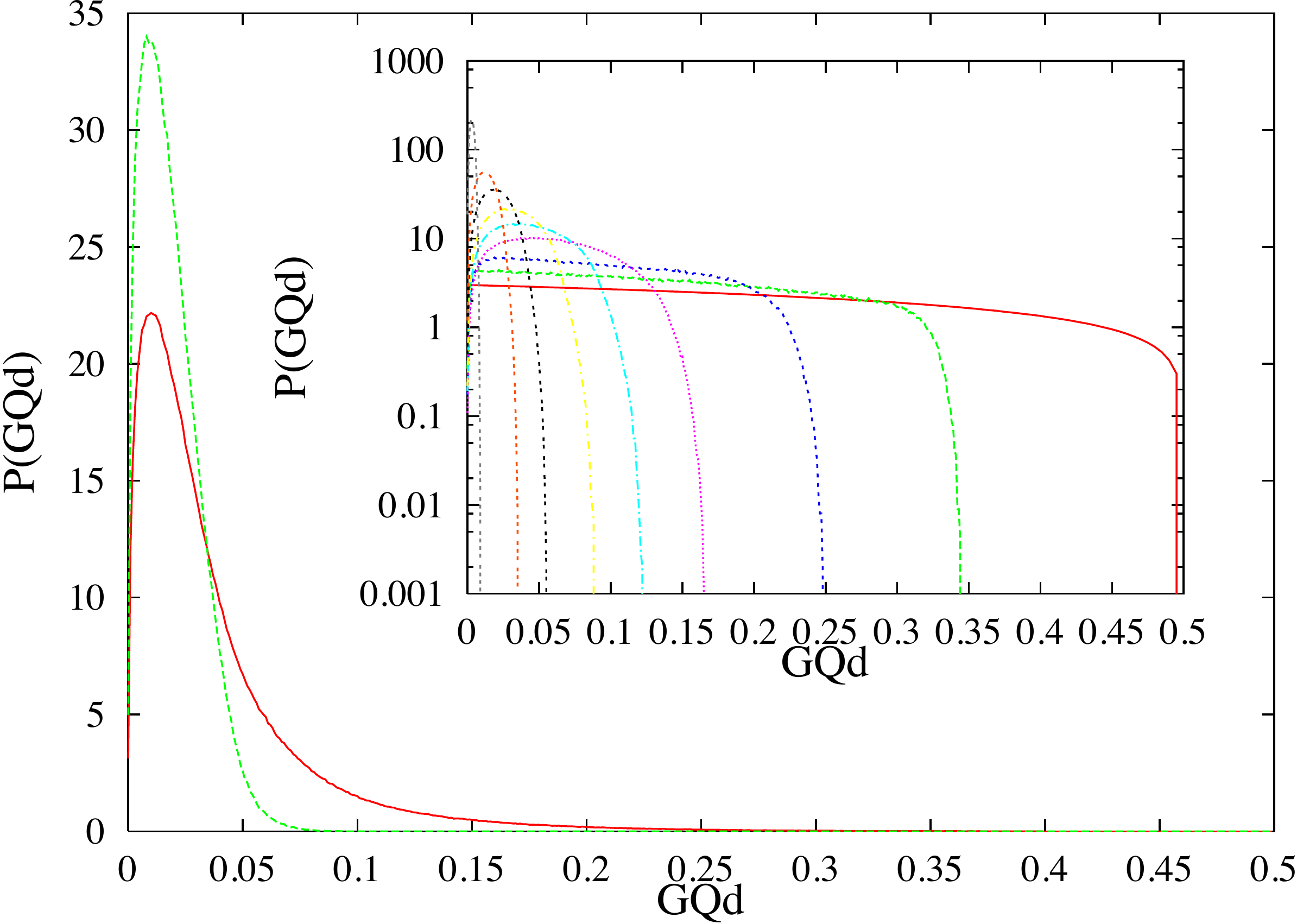}
\caption{(Color online) Plot of the probability (density) distribution of finding a state
$\rho$, pure or mixed, of two qubits with a given value of the discord GQd. The lower
curve corresponds to all states, while the upper curve depicts the same distribution for
separable states only (PPT states). Notice the strong bias of both curves towards low
values of GQd. The inset depicts similar probability (density) distributions for states
with a particular value of the participation ratio only. As we increase the value of R
(values for R=1, 1.3, 1.6, 2, 2.3, 2.6, 3, 3.3 and 3.8, from right to left), the range of
available GQd's diminishes. The distribution for pure states (R=1) generated according to
the Haar measure is analytic. See text details.}
\label{fig2}
\end{center}
\end{figure}

How are finite $D-$states distributed in Hilbert space? We refer
the reader to  Fig. 2, that depicts the probability density
distribution of finding a given value $D$ for the geometric
distance GQd of reference \cite{geom}  in the whole space of two
qubits. We plot the probability (density) distribution of finding
a two qubits-state $\rho$, pure or mixed,  with a given value $D$
of the distance GQd. The lower curve corresponds to all states,
while the upper curve depicts the same distribution for separable
states only (PPT states). Notice the strong bias of both curves
towards low values of GQd. The inset depicts similar probability
(density) distributions, but for states with a particular value of
the participation ratio $R$ only. As we increase the value of $R$
(we show values for R=1, 1.3, 1.6, 2, 2.3, 2.6, 3, 3.3 and 3.8,
from right to left), the range of available GQd's diminishes, as
expected. The distribution for pure states ($R=1$) generated
according to the Haar measure can be obtained in analytic fashion
as follows. One starts from the general pure state form (Schmidt
decomposition)

\be \label{cuatro} \cos{\theta} |00\rangle + \sin{\theta}
|11\rangle,\ee

\noindent
and compute $Qd(\theta)=(1/2)\sin^2{2\theta}$. For
instance, if we have a concurrence $C=\sin{2\theta}$, the
state-distribution is known to be $P(C^2)=(3/2)\sqrt{(1-C^2)}$.
Thus, one easily ascertains that

\be \label{cinco} P(Qd)=3\sqrt{(1-2Qd)},\ee which coincides with
the numerical result.

\begin{figure}[htbp]
\begin{center}
\includegraphics[width=8cm]{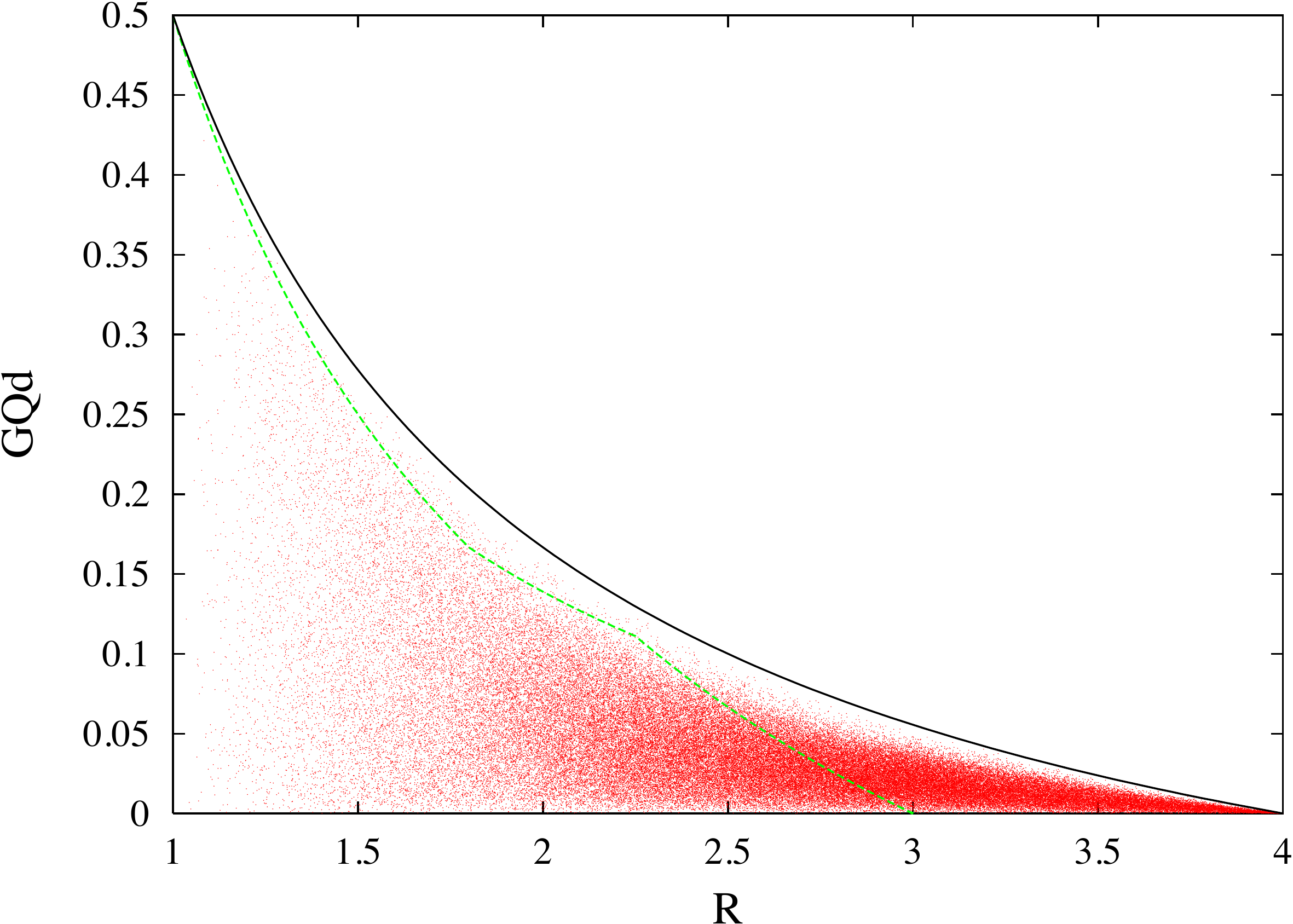}
\caption{(Color online) Plot of the GQd vs the participation ratio R for a sample of $10^6$
mixed states of two qubits. The lower curve corresponds to MEMS states, while the solid
line corresponds to the maximum GQd value compatible with a given R. This maximum is
attained by Werner states for the measure GQd of quantum discord. See text for details.}
\label{fig3}
\end{center}
\end{figure}

\begin{figure}[htbp]
\begin{center}
\includegraphics[width=8cm]{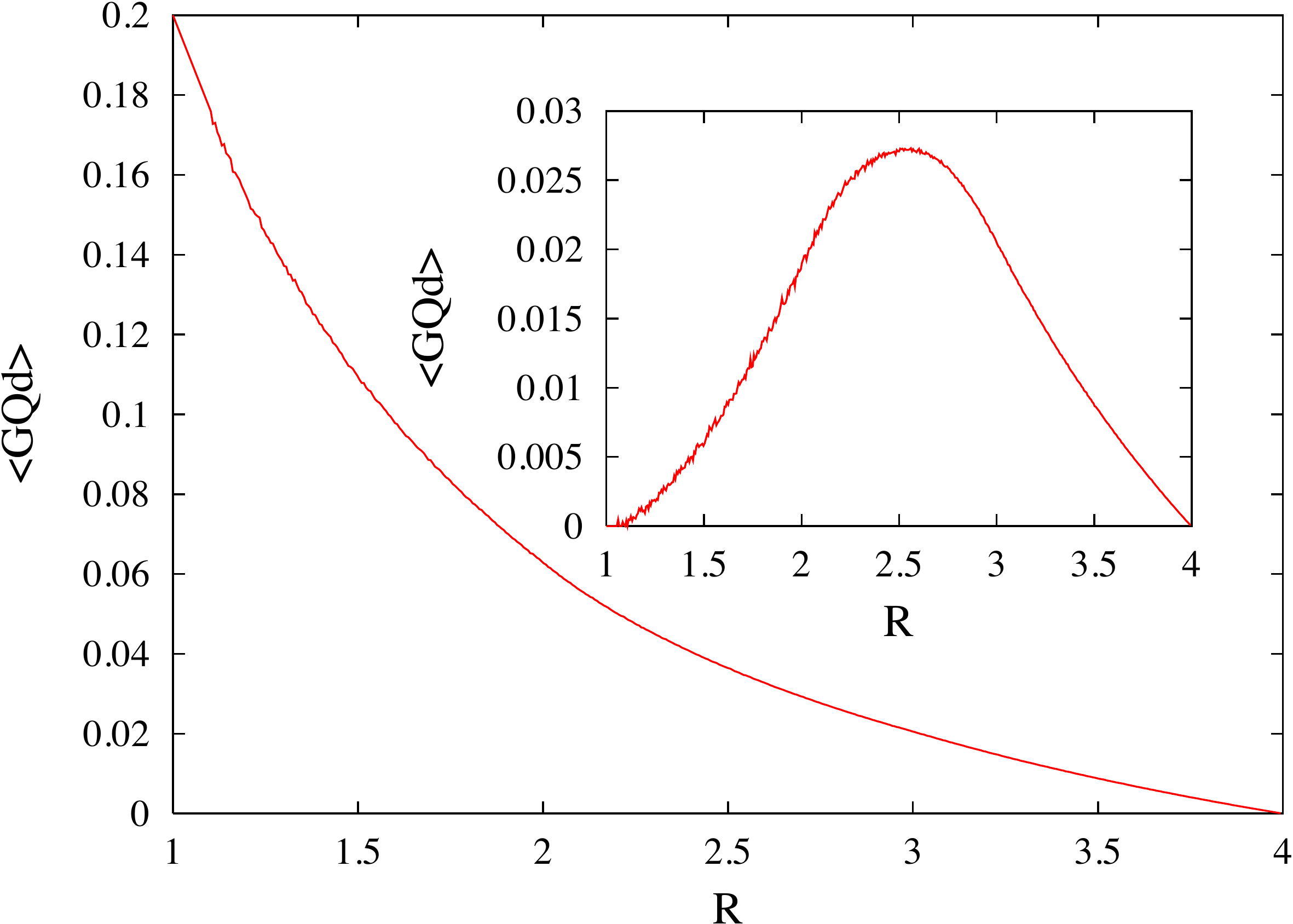}
\caption{(Color online) Plot of the mean value of the quantum discord measure GQd vs the
participation ratio R for all the space of two qubit states. The monotonic behavior of
GQd is apparent. The value for pure states is analytic ($\frac{1}{5}$).
The inset depicts the same quantity GQd only for separable states. Notice the expected
behavior for the quantum discord (null for separable and maximally mixed states). Both
curves coincide in the range R $\in [3,4]$ for all states are separable. See text for
details.}
\label{fig4}
\end{center}
\end{figure}

In Fig. 3 we see a plot of the geometric distance GQd vs. $R$ for
a sample of $10^6$ mixed states of two qubits. The lower curve
corresponds to MEMS, while the solid line corresponds to the
maximum GQd value compatible with a given $R$. Remarkably enough,
we encounter that this maximum is attained by Werner states. That
is, these states maximize the geometric measure of Ref.
\cite{geom}. This graph constitutes  a nice illustration of the
fact that zero-discord is a rather rare event.

%%%%%%%%%%%%%%%%%%%%%%%%%%%%%%%%%%%%%%%%%%%%%%%%%%%%%%%%%%%%%%%%%%%%%%%%%%%%%%%%%%%%%%%%%%%%%%%%%%%%%%%%%%
%%%%%%%%%%%%%%%%%%%%%%%%%%%%%%%%%%%%%%%%%%%%%%%%%%%%%%%%%%%%%%%%%%%%%%%%%%%%%%%%%%%%%%%%%%%%%%%%%%%%%%%%%%
Let us delve into the case of Werner states. Since we seek maximum
correlations (quantal + classical), we shall consider states which
are diagonal in the Bell basis, since nonlocal correlations tend
to concentrate after some depolarizing process \cite{depol}. If we
do so, states (\ref{rhoBloch}) should possess null values for
${\bf x}$ and ${\bf y}$. In other words, we have Bell diagonal
states. We consider the paradigmatic case of Werner states of the
type diag$(1-3x,x,x,x)$ ($x\in[0,\frac{1}{4}]$) in the Bell basis
$\{
|\Phi^{+}\rangle,|\Phi^{-}\rangle,|\Psi^{+}\rangle,|\Psi^{-}\rangle
\}$. These are a special case of the ones considered in Ref.
\cite{geom} (states of maximally mixed marginals with $T_{11}=0,
T_{33}=1-4x, T_{22}=-T_{33}$). Thus,  we have $D=\frac{1}{2}
(1-4x)^2=\frac{1}{6} \bigg( \frac{4}{R}-1 \bigg)$, which is
optimal.

We pass now to the mean value of the geometric measure GQd as a
function of the participation ratio $R$ for all the space of two
qubits (Fig. 4). The monotonic behavior of the GQd is apparent.
The value for pure states is analytic ($\frac{1}{5}$). The inset
depicts the same quantity GQd, but only for separable states.
Notice the expected behavior for the quantum discord (null for
separable and maximally mixed states). Both curves coincide in the
range $R \in [3,4]$.

Since we encounter the Werner states to be most ``discordant''
ones for a given value of the participation ratio $R$, we may
wonder why this is not the case for the original quantum discord
measure Qd of \cite{olli}. In point of fact, states
diag$(1-3x,x,x,x)$, with $x\in[\frac{1}{4},\frac{1}{3}]$) happen
to be of maximal Qd in the region $R\in[3,4]$, where all states
are separable! This implies that a depolarizing process would
concentrate quantum correlations for Qd only where just classical
correlations exist, which seems to be a contradiction. Therefore,
we are forced to abandon the initial assumption that a state
$\rho$, who has been through some depolarizing channel,
concentrates quantum discord. This fact is sustained by the
evidence that Bell diagonal states do not optimize Qd for a given
$R$. In point of fact, states that concentrate entanglement for a
given $R$, that is, MEMS states, possess the form

\begin{equation} \label{MEMSBell}
\left( \begin{array}{cccc}
g(x)+\frac{x}{2} & 0 & 0 & 0\\
0 & g(x)-\frac{x}{2} & 0 & 0\\
0 & 0 & \frac{1-2g(x)}{2} & \frac{1-2g(x)}{2}\\
0 & 0 & \frac{1-2g(x)}{2} & \frac{1-2g(x)}{2}
\end{array} \right),
\end{equation}

\noindent in the Bell basis, with $g(x)=1/3$ for $0\le x \le 2/3$,
and  $g(x)=x/2$ for $2/3 \le x \le 1$ (the quantity $x$ being
equal to the concurrence $C$). Notice the aforementioned tendency
of states to concentrate correlations (here entirely  from an
entanglement origin) in view of the quasi diagonal form of
(\ref{MEMSBell}). In addition, MEMS maximize Qd in the R-range
$[1, 1.8 ]$. Thus, maximum discord and maximum entanglement
coincide for low-purity values.
%%%%%%%%%%%%%%%%%%%%%%%%%%%%%%%%%%%%%%%%%%%%%%%%%%%%%%%%%%%%%%%%%%%%%%%%%%%%%%%%%%%%%%%%%%%%%%%%%%%%%%%%%%
%%%%%%%%%%%%%%%%%%%%%%%%%%%%%%%%%%%%%%%%%%%%%%%%%%%%%%%%%%%%%%%%%%%%%%%%%%%%%%%%%%%%%%%%%%%%%%%%%%%%%%%%%%

Fig. 2 should be now compared to Fig. 5, that gives, in the
fashion of Zambrini et al. \cite{zambrini}, the probability
(density) distribution for the true discord  Qd for all
two-qubits' states (lower curve). The  more peaked curve
corresponds  to separable states only. The inset refers to   a
more original scenario and depicts the density distribution of the
classical correlations (CC). Classical and quantum correlations
are distributed in rather similar fashion because most states
exhibit very small $D$'s.\\

We pass now to Fig. 6. It is a sample plot for Qd vs. $R$ for
$10^6$ mixed states  of two qubits. The upper dashed curve
corresponds to MEMS states, while the solid one is that for Werner
states. The inset depicts the mean value of Qd vs $R$. Notice the
change in the curve for all states being separable ($R \geq$ 3).
The horizontal line represents the analytic value for pure states
generated according to  the Haar measure ($\frac{1}{3\log 2}$).

\begin{figure}[htbp]
\begin{center}
\includegraphics[width=8cm]{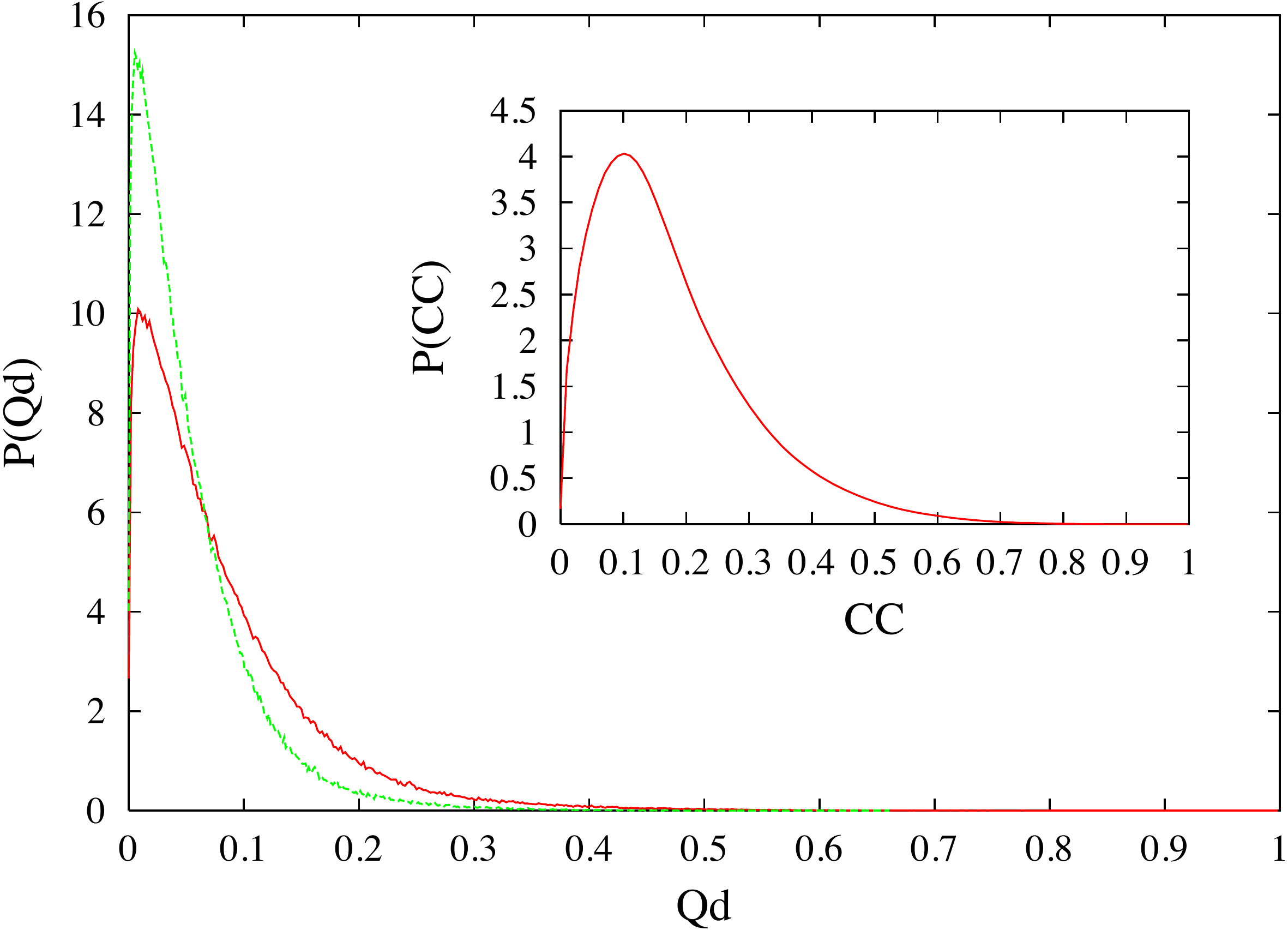}
\caption{(Color online) Probability (density) distribution for discord Qd for all states
of two qubits (lower curve). The lower one (more peaked) corresponds to separable states
only. The inset depicts a similar distribution for the classical correlations CC. See text for
details.}
\label{fig5}
\end{center}
\end{figure}

\begin{figure}[htbp]
\begin{center}
\includegraphics[width=8cm]{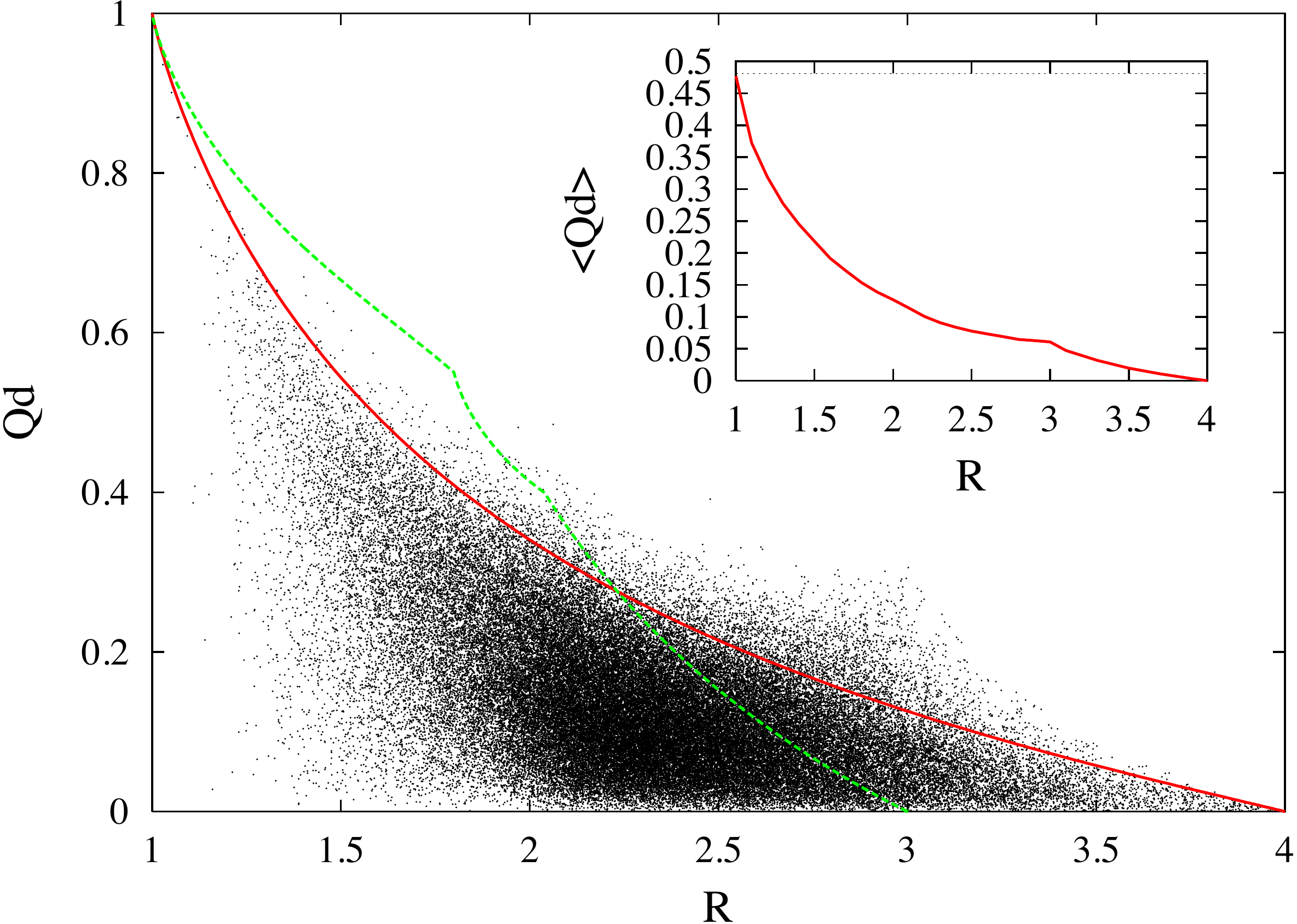}
\caption{(Color online) Sample plot for Qd vs R for $10^6$ mixed states of two qubits. The
upper dashed curve corresponds to MEMS states, while the solid one to Werner states. The inset depicts
the mean value of Qd vs R. Notice the change in the curve for all states being separable (R $\geq$ 3). The
horizontal line represents the analytic value for pure states generated according to the Haar measure
($\frac{1}{3\log 2}$). See text for details.}
\label{fig6}
\end{center}
\end{figure}

\section{Bell inequalities and geometric measure for quantum discord}

Most of our knowledge on Bell inequalities and their quantum
mechanical violation is based on the CHSH inequality
\cite{NC00,GisinTheorem}. With two dichotomic observables per
party, it is the simplest nontrivial Bell inequality for the
bipartite case with binary inputs and outcomes. Quantum
mechanically, these observables reduce to ${\bf A_j}({\bf
B_j})=\bf{a_j}(\bf{b_j}) \cdot \bf{\sigma}$, where
$\bf{a_j}(\bf{b_j})$ are unit vectors in $\mathbb{R}^3$ and
$\bf{\sigma}=(\sigma_x,\sigma_y,\sigma_z)$ the Pauli matrices.
Violation of CHSH inequality requires
% $Tr[\rho_{ij}^{(R)} B_{CHSH}]$, that is,
the expectation value of the operator
$B_{CHSH}=
{\bf A_1}\otimes {\bf B_1} + {\bf A_1}\otimes {\bf B_2} + {\bf
A_2}\otimes {\bf B_1}  -  {\bf A_2}\otimes {\bf B_2} $
to be greater than two. In this vein one should make
reference to the Tsirelson-bound, also known as Tsirelson's
inequality. It is indeed an inequality that imposes an upper limit
to quantum mechanical correlations between distant events. It
relates to the  discussion and experimental determination of
whether local hidden variables are required for, or even
compatible with, the representation of experimental results. This
is of  particular relevance to  EPR's thought experiment and to
the CHSH inequality. It is named for B. S. Tsirelson, who derived
it \cite{tire}.\\

Let us now consider then the case of the maximum violation of a
Bell inequality, in the form of the CHSH Bell inequality for two
qubits.  Fig. 7 depicts the nonlocality measure $B_{CHSH}^{\max}$
vs $GQd$ for a random sample of uniformly generated two
qubit-mixed states. The horizontal curve at the height ``two"
represents the limit for local variable model theories (LVM) to
hold, whereas the upper one at $2\sqrt{2}$ represents the
Tsirelson-bound for quantum mechanics. As we can appreciate, each
quantum state lies between the two curves, and the general trend
is a certain correlation between both quantities. This remarkable
behavior can be explained as follows. Since we seek maximum
violation of the CHSH inequality, those states necessarily must
concentrate their nonlocal correlations. Therefore, a considerable
subset of states that fills the entire region between these curves
is that of Bell diagonal states. In point of fact, those Bell
diagonal states that are less nolocal are precisely the previously
discussed Werner states, that is, diag$(1-3x,x,x,x)$
($x\in[0,\frac{1}{4}]$), which exhibit a maximum amount of GQd.
Therefore, when calculating their nonlocality measure, we find it
to be \cite{batalla}(J. Batle and M. Casas, e-print
arXiv:quant-ph/1102.4653.) $B_{W}=2\sqrt{2}(1-4x)$, which implies
that the lower curve in Fig. 7 is of the  form
$B_{CHSH}^{\max}=4\sqrt{GQd}$.\\

Now, what is the nature of the upper curve in Fig. 7? As explained
in Ref. \cite{batalla} there is a limit to those states that
maximize the CHSH inequality. These states are given in the form
in Eq. (17) of \cite{batalla}, namely, ${\rm diag}(1-x,x,0,0)$
($x\in[0,\frac{1}{2}]$) also in the Bell basis. These maximally
nonlocal mixed states (MNMS) maximally violate the CHSH inequality
for a given value of the participation ratio $R$. Their
nonlocality is given by $B=2\sqrt{2}\sqrt{(1-x)^2+x^2}$. Obtaining
their concomitant GQd measure in the usual manner, we obtain
$Gqd=\frac{1}{2}(1-2x)^2$. Combining both relations, we naturally
obtain $B_{CHSH}^{\max}=2\sqrt{1-2GQd}$.

\begin{figure}
\begin{center}
\includegraphics[width=8.6cm]{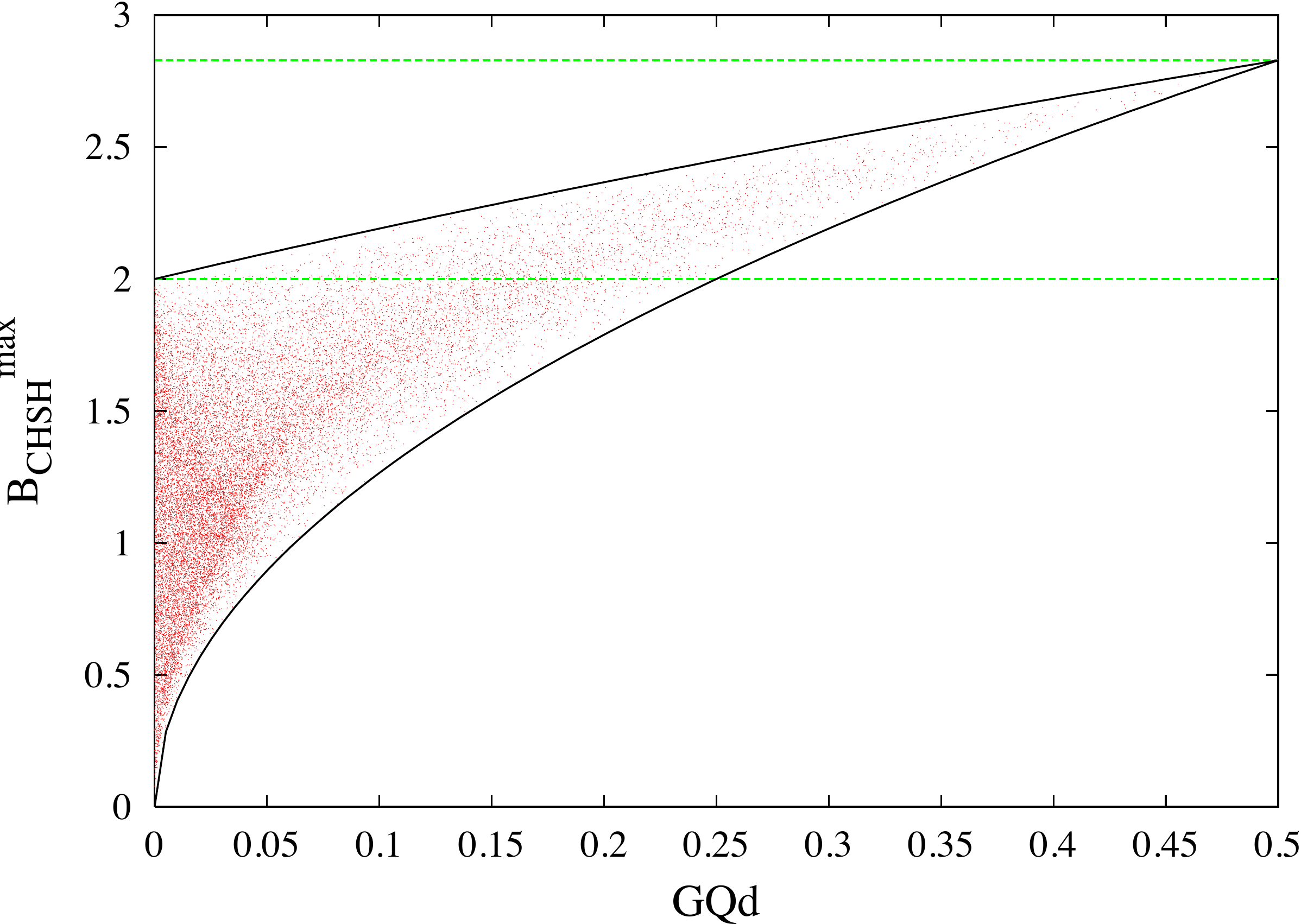}
\caption{(Color online) Sample plot of two qubit mixed states for
the quantities $B_{CHSH}^{\max}$ vs $GQd$. The two curves are
obtained in analytical fashion. See text for details.}
\label{fig7}
\end{center}
\end{figure}

\section{Conclusions}

We have performed a systematic survey of the two qubits
Hilbert's space so as to assess the main features of the geometric
measure of quantum discord  $D$ advanced in \cite{geom}. We
have shown that, when considering the typical behavior of general (pure or mixed)
states of two qubits, the geometric measure of quantum discord is strongly
correlated with the measure of quantum discord originally advanced by
Zurek and Ollivier. We investigated the connection between quantum
discord and degree of mixedness as measured by the participation ratio $R$.
As a general trend we observed that as $R$ increases the range of possible
values of the discord decreases, only small vales of $D$ becoming available.
We showed that the maximum values of $D$ compatible with given values of $R$
are attained by the Werner states. The behaviour of MEMS in connection with
quantum discord and degree of mixedness was also addressed. Finally, we examined
the connection between the geometric measure of quantum discord and
non-locality (as measured by the maximum violation of a bell inequality
within the CHSH scenario). Our results indicate that there exists a clear
tendency of non-locality to increase as one considers two-qubits states
exhibiting increasing values of quantum discord.

\noindent{\bf acknowledgments-} This work was partially supported
by project FIS2008-00781/FIS (MICINN) and FEDER (EU), by the programs
FQM-2445 and FQM-207 of the Junta de Andalucia-Spain, and by CONICET
(Argentine Agency).

\end{document}